# Gain of squeezing via photon subtractions

Mikhail S. Podoshvedov[1,2,3], Sergey A. Podoshvedov[1,2*] and Sergei P. Kulik[2,4]

[1]*Laboratory of quantum information processing and quantum computing, South Ural State University (SUSU), Lenin Av. 76, Chelyabinsk, Russia*
[2]*Laboratory of quantum engineering of light, South Ural State University (SUSU), Lenin Av. 76, Chelyabinsk, Russia*
[3]*Kazan Quantum Center, Kazan National Research Technical University named after A.N. Tupolev, Kazan, Russia*
[4]*Quantum Technology Centre, M.V. Lomonosov Moscow State University, Moscow, Russia*

[*]sapodo68@gmail.com

**Abstract:** We develop a method for generating a more squeezed than single-mode squeezed vacuum (SMSV) state by subtracting 2,4,6 photons from it. In general, the more photons are subtracted, the more gain of the squeezing (more of $3\ dB$) is observed in the measurement-induced continuous variable (CV) states of definite parity. However, the two-photon subtraction strategy is practically preferred. It can be implemented with higher success probability, wider squeezing gain width $\sim 5\ dB$ and least quadrature variance in the corresponding range of initial squeezing. We demonstrate the mitigating effect of a photon-number resolving (PNR) detector with non-unit quantum efficiency on the output characteristics of the measurement-induced CV states, resulting in their slight decrease compared to ideal photon subtraction. Use of a single photon in addition to the SMSV state at beam splitter (BS) input with subsequent registration of odd number of photons (say, 3 photons) allows the implementation of the measurement-induced even CV state which is several times brighter than the initial state and has lower quadrature noise.

## 1. Introduction

Squeezed states of light have oddity that one of the quadrature variance is less than that of vacuum state [1-4]. The definition of squeezed light is related to the squeezing of noise in one of the quadrature components. This is due to the fact that the other quadrature component has variance above the variance of the vacuum in order to satisfy Heisenberg uncertainty principle [5]. Typical examples of the squeezed state are the Gaussian states such as SMSV and related with it two-mode squeezed vacuum (TMSV) states which, from a formal point of view, are implemented by application of the corresponding squeezing operators [6,7]. SMSV state is obtained when a powerful laser field propagates through a second-order nonlinear optical medium splitting into photons of lower energy in an indistinguishable manner, that is, generated photons are indistinguishable in all their parameters: frequency, polarization and direction of propagation [8]. In addition to its fundamental importance of the effect, observation of the squeezing contributed to the subsequent development of optical homodyne tomography [9-11] resulted in techniques of creating, manipulating and measuring quantum states of light achieving $9\ dB$ squeezing in [12] and $10\ dB$ squeezing in [13]. Note that more squeezed states at some near unfrared wavelengths are demonstrated in [14,15].

Squeezed light is useful in ultra-precise estimate of unknown optical phase [16,17], where the precision of the interferometer is determined by the phase uncertainty of the states used. The idea is used in incorporating squeezed light into gravitational waves detectors to enhance their sensitivity [18-20]. Review of squeezed states with a focus on experimental observations is presented in [21]. Another application of squeezed light involves quantum teleportation of an unknown qubit with TMSV communication resource and the resulting fidelity exceeded the classical benchmark [22]. Thereafter, numerous efforts have been made to modify the protocol for the teleportation of other optical states [23,24]. Squeezed states of light, both SMSV and TMSV, are the cornerstone of quantum engineering of light, extending from heralded generation of single photons when one channel of the entangled state is monitored by a single-photon detector [25,26] to various modifications of conditional generation of optical Schrödinger cat states (SCSs) [27-29].

The generation of highly squeezed states can be quite challenging [15], therefore methods that allow increasing the squeezing of initially weakly squeezed light are of interest. Photon subtracted and photon added SMSV states are among those CV states that are experimentally realizable [30-32]. The theory of subtraction of $k-$photons from the initial Gaussian state $\rho$ is based on application of the annihilation operator to the appropriate power $k$ to the initial state, i.e. $\rho_{-k} \sim a^k \rho a^{+k}$ which from a physical point of view can be approximated by passing the SMSV state through a highly transmitting BS with subsequent registration of $k$ photons in the adjacent measuring mode [33,34]. However, the annihilation and creation operators are non-unitary and cannot be directly implemented in the laboratory. A realistic scenario with variable BS parameter leads to the generation of the hybrid entangled state of light [35] and BS transmissivity is part of the measurement-induced CV states [36]. In general, the practical approach associated with generation of the hybrid entangled state after the passage of the SMSV state through an arbitrary BS deserves more attention [37] taking into account the progress in development of PNR technology [38,39]. It provides the opportunity to take into account the redistribution of input photons in various regimes of the BS which can lead to a significant change in the nonclassical properties of the generated CV states [40], in particular, to increase in their brightness [41]. Enhancement of the squeezing of SMSV state by subtracting two photons is demonstrated in [42]. The hybrid entanglement can be realized by subtracting photons from initial Gaussian state [43,44] even for any measurement outcome [45].

Here we develop a practical approach to implementation of squeezing-enhanced states, that is, those whose quadrature noise is less than that of the original SMSV state, via subtraction of even number of photons. Optimization by the beam splitter parameter allows to estimate the maximum gain of the squeezing of the measurement-induced CV states of definite parity as well as the best strategy for the number of photons subtracted in terms of minimal quadrature variance, increased success probability and squeezing gain width. The output characteristics of the measurement-induced CV states turned out to be resistant to PNR detector imperfections, i.e. photon losses can only slightly reduce them. Expansion of the photon subtraction problem by adding a single photon allows one to observe more squeezed and brighter measurement-induced CV states of a certain parity.

## 2. Photon subtraction as a way to enhance squeezing

Let's consider the initial SMSV state

$$|SMSV(y)\rangle = \frac{1}{\sqrt{\cosh s}} \sum_{n=0}^{\infty} \frac{y^n}{\sqrt{(2n)!}} \frac{(2n)!}{n!} |2n\rangle, \qquad (1)$$

the canonical form of which [3] can be represented in the case of replacing the squeezing parameter $y$ on the squeezing amplitude $s > 0$, that is $y = \tanh s/2$, which provides the range of its change $0 \leq y \leq 0.5$. In addition, the squeezing in decibels $S = -10 \lg(\exp(-2s)) \, dB$ and the mean number of photons $\langle n_{SMSV} \rangle = \sinh^2 s$ can be used for the description. In addition to the parameters, one can also consider its squeezed property, which is expressed in reducing the noise of one of the quadrature components: $X_1 = (a + a^+)/2$ and $X_2 = (a - a^+)/2i$ below vacuum noise, i.e.

$$S_{SMSV} = \langle X_2^2 \rangle_{SMSV} = \frac{exp(-2s)}{4}, \tag{2}$$

as $\langle X_1 \rangle_{SMSV} = \langle X_2 \rangle_{SMSV} = 0$. The state (1) is a minimum uncertainty state with Gaussian Wigner function [21]. To try to surpass the squeezing of the Gaussian state, it is necessary to use an operation that converts it to a non-Gaussian, which in practice results in the use of photon subtraction procedure.

The photon subtraction can be realized with the help of PNR detector [39] used in the measuring mode of the BS through which the initial SMSV state passes [35,37]. As shown in the Appendix A, the photon subtraction by the ideal PNR detector guarantees the generation of one of the possible CV states of definite parity depending on the number of measured photons in the measuring mode either even

$$|\Psi_{2m}^{(0)}(y_1)\rangle = \frac{1}{\sqrt{Z^{(2m)}(y_1)}} \sum_{n=0}^{\infty} \frac{y_1^n}{\sqrt{(2n)!}} \frac{(2(n+m))!}{(n+m)!} |2n\rangle, \tag{3}$$

or odd

$$|\Psi_{2m+1}^{(0)}(y_1)\rangle = \sqrt{\frac{y_1}{Z^{(2m+1)}(y_1)}} \sum_{n=0}^{\infty} \frac{y_1^n}{\sqrt{(2n+1)!}} \frac{(2(n+m+1))!}{(n+m+1)!} |2n+1\rangle, \tag{4}$$

in dependency on parity of the measurement outcomes either $k = 2m$ or $k = 2m + 1$. The subscript $2m, 2m + 1$ is responsible for the number of subtracted photons, while the superscript accounts for the number of additional input photons in auxiliary mode. In considered case, $(0)$ corresponds to the input vacuum state in auxiliary mode. The notations used here $y_1$, $Z^{(2m)}(y_1)$ and $Z^{(2m+1)}(y_1)$ and also the beam splitter parameter $B$ are described in more detail in the Appendix A. As follows from the consequences presented in the Appendix A, the amplitudes $c_k^{(0)}(y_1, B)$ of the hybrid entangled state in equation (A1) determine the probability distribution of generating measurement-induced CV states of certain parity

$$P_k^{(0)}(y_1, B) = \frac{c_k^{(0)2}(y_1,B) Z^{(k)}(y_1)}{\cosh s} = \frac{\sqrt{1-4y_1^2(1+B)^2}(y_1 B)^k Z^{(k)}(y_1)}{k!}. \tag{5}$$

Quadrature variance of the measurement-induced CV state becomes

$$(\Delta X_2^2)_k^{(0)} = \frac{1}{4} + \frac{\langle n_k^{(0)} \rangle}{2}(1 - 2y_1) - y_1(k+1), \tag{6}$$

where the mean number of photons is given by [41]

$$\langle n_k^{(0)} \rangle = y_1 \frac{Z^{(k+1)}(y_1)}{Z^{(k)}(y_1)}. \tag{7}$$

An increase in the squeezing compared to the initial SMSV state is observed only in the case of an even number of subtracted photons and, as a consequence, for even measurement-induced CV states in equation (3). The lack of the squeezing gain in odd CV states can be explained by considering the case of $s \to 0$, when the state in Eq. (4) approaches the single photon whose less noisy quadrature is equal to $(\Delta X_2^2)_{|1\rangle} = 3/4$. On the contrary, even measurement outcomes generate even CV states that are close enough to vacuum in the case of $s \to 0$ that provides $(\Delta X_2^2)_k^{(0)} \approx (\Delta X_2^2)_{|0\rangle} = 1/4$. The discrepancy in quadrature uncertainties between even and odd measurement-induced CV states in the case of $s \to 0$ leads to a difference in their behavior with $s$ increasing, in particular, to the fact that the quadrature component of the odd CV states is noisier compared to the original SMSV state.

Unlike odd CV states, even ones can exhibit less quadrature noise compared to the initial SMSV state. Indeed, the graphs presented in figure 1(a) confirm the fact. Optimized by the parameter $B$, which means choosing the minimum value of the quadrature uncertainty in Eq. (6) by varying the parameter $B$ for a given $S$, dependences of the quadrature squeezing $S_k^{(0)}$ on the initial squeezing $S$ are presented in Figure 1(a) for $k = 2,4,6$. For comparison, the dependence of the less noisy quadrature component of the SMSV state $S_{SMSV}$ in Eq. (2) is also shown in black in Fig. 1(a). We can talk about the generation of less quadrature-noisy CV state in the case of $S_{SMSV} > S_k^{(0)}$. In the opposite case of $S_k^{(0)} \geq S_{SMSV}$, SMSV state is the less quadrature-noisy CV one. Almost horizontal sections in $S_2^{(0)}, S_4^{(0)}, S_6^{(0)}$ dependencies are observed which can be related to the behavior of optimizing $B_{opt,k}^{(0)}$ which, as shown in Fig. 1(b), starts with $B_{opt,k}^{(0)} < 1$, that is, with highly transmitting BSs with $T = t^2 \gg R = r^2$, increases almost linearly upwards $B_{opt,k}^{(0)} > 1$ after some value of $S$, thus, switching to more reflective beam splitters with $R > T$. Although, it is worth noting that in the case of $k = 2$, the parameter $B_{opt,k}^{(0)}$ continues to remain less of one in the selected range of change of the initial $S$, i.e., $B_{opt,k}^{(0)} < 1$, without reaching the level of $B_{opt,k}^{(0)} = 1$ which corresponds to the balanced beam splitter.

To quantify the gain of the squeezing of the SMSV state from which photons are subtracted, we show its dependence $g_k^{(0)} = -10 \cdot lg\left(S_k^{(0)}/S_{SMSV}\right)$ expressed in decibels ($dB$) on the initial squeezing $S$ in Figure 1(c). Actually, the range of $S$, where $g_k^{(0)} > 0$ can be taken as the width of the gain squeezing realized by photon subtraction. Negative values of $g_k^{(0)} < 0$ indicate that the quadrature noise of the measurement-induced CV states become larger than the one of the SMSV state. This is due to the fact that $S_{SMSV}$ can decrease faster than $S_k^{(0)}$ with $S$ growing as shown in figure 1(a). In figure 1(d) we show the success probabilities $P_k^{(0)}$ following from equation (5) as function of the squeezing parameter $S$. The probabilities, especially $P_2^{(0)}$, demonstrate significant gain with input squeezing $S$ increasing. The parameters $S$ and $B_{opt,k}^{(0)}$ that provide the maximum squeezing gain $g_{k,max}^{(0)}$ and corresponding them the probabilities of success $P_k^{(0)}$ are presented in Table 1.

Overall, subtraction of two photons is the best strategy for the squeezing enhancement despite $g_{6,max}^{(0)} > g_{4,max}^{(0)} > g_{2,max}^{(0)}$. The main advantage of the two-photon subtraction strategy is its fairly high success probability $P_2^{(0)}$ compared to 4 and 6 photons subtraction cases. Another advantage is wider squeezing gain width $\sim 5\ dB$ compared to widths related with $k = 4, 6$ subtracted photons. Moreover, the absolute values of $S_2^{(0)}$ are less than $S_4^{(0)}, S_6^{(0)}$ in wide range of change of $S$, i.e. for $5\ dB > S > 1.7\ dB$. In fact, in $1.7\ dB > S$ range, subtraction of 4 and 6 photons can become more attractive for the gain of the squeezing. The facts can only increase the practical significance of the procedure of subtracting two photons from the initial SMSV state to achieve gain of the squeezing. Note that decrease of quadrature noise is also observed at other values of $B$ different from $B_{opt,k}^{(0)}$, i.e. for $B \neq B_{opt,k}^{(0)}$, but by a smaller amount. But this can also be beneficial in terms of increasing the probability of success.

|  | $k = 2$ | $k = 4$ | $k = 6$ |
|---|---|---|---|
| $S, dB$ | 2.026 | 1.159 | 0.841 |
| $B_{opt,k}^{(0)}$ | $0.02\ (t = 0.99)$ | $0.02\ (t = 0.99)$ | $0.02\ (t = 0.99)$ |
| $g_{k,max}^{(0)}, dB$ | 2.551 | 2.952 | 3.119 |
| $P_k^{(0)}$ | $1.267 \cdot 10^{-5}$ | $2.213 \cdot 10^{-11}$ | $1.92 \cdot 10^{-17}$ |

**Table 1.** The values $S, dB$ and $B_{opt,k}^{(0)}$ presented provide the maximum squeezing gain $g_{k,max}^{(0)}, dB$ of the SMSV state from which 2,4,6 photons, respectively, are subtracted. The corresponding numerical values of $P_k^{(0)}$ are also shown.

## 3. Adding photon to generate brighter squeezed state

As shown above, more efficient photon subtraction in the terms of the gain squeezing can be implemented just in the practical range of initial squeezing up to $S < 5\ dB$ when two photons are subtracted. Generation of the SMSV state with $S > 5\ dB$ no longer guarantees the gain squeezing through the photon subtraction procedure. The squeezed state with $S < 5\ dB$ can be considered as CV state with a predominance of the vacuum state which can only reduce the probability of photon subtraction. To increase the probability of the multiphoton outcomes, let us consider the procedure of adding single photon to the initial SMSV state [33] followed by the subtraction of $k$ photons in order to generate new non-Gaussian CV states different from those considered above. Then, we can talk about the combined action of adding and subtracting photons.

As follows from the results of the Appendix A, in the case of an additional input single photon, the corresponding quadrature variance of the measurement-induced CV states is calculated using the following formulas

$$(\Delta X_2^2)_0^{(1)} = \frac{1}{4} + \frac{1}{2}\langle n_0^{(1)}\rangle(1 - 2y_1) - 2y_1, \tag{8}$$

for $k = 0$ and

$$(\Delta X_2^2)_k^{(1)} = \frac{1}{4} + \frac{1}{2}\langle n_k^{(1)}\rangle(1 - 2y_1) - ky_1 + \frac{2y_1 B}{k}\frac{R_k^{(1)}(y_1, B)}{G_k^{(1)}(y_1, B)}, \tag{9}$$

for $k > 0$, where the mean number of photons in SMSV state to which one photon is added and $k$ photons are subtracted is given by

$$\langle n_k^{(1)}\rangle = \frac{y_1 \frac{d}{dy_1} G_k^{(1)}(y_1, B)}{G_k^{(1)}(y_1, B)}. \tag{10}$$

Here all the functions $G_k^{(1)}(y_1, B)$ and $R_k^{(1)}(y_1, B)$ used as well as the measurement-induced CV states realized by adding single photon and subtracting $k$ photons are presented in the Appendix A. Note that the probabilities of realizing the measurement-induced CV states of a certain parity in equations (A4-A6) are given by the following distribution

$$P_k^{(1)}(y_1, B) = \frac{\sqrt{1 - 4y_1^2(1+B)^2}}{1+B} \begin{cases} \frac{B}{y_1} G_0^{(1)}(y_1, B), & \text{if } k = 0 \\ \frac{(y_1 B)^{k-1}}{k!} k^2 G_k^{(1)}(y_1, B), & \text{if } k \neq 0 \end{cases}. \tag{11}$$

In figure 2(a) we show the dependences of the squeezing $S_k^{(1)}$ of the measurement-induced CV states with single photon added and $k = 1,2,3,4,5,6$ photons subtracted. Here, as above, optimization by $B$ means finding the minimum value of quadrature variance $(\Delta X_2^2)_k^{(1)}$ in equation (9) by varying $B$ for given value of initial squeezing $S$. The optimizing values $B_{opt,k}^{(1)}$ in figure 2(b) take values less than 1, that is, $B_{opt,k}^{(1)} < 1$ which indicates in favor of using highly transmitting BS with $T \gg R$. Almost all $B_{opt,k}^{(1)}$ take on the same value, that is, $B_{opt,1}^{(1)} \approx B_{opt,2}^{(1)} \approx B_{opt,4}^{(1)} \approx B_{opt,6}^{(1)}$, except for the functions $B_{opt,3}^{(1)}$ and $B_{opt,5}^{(1)}$, which are different from the rest and have the bell shapes on a certain section of $S$, however, remaining less than 1 which indicates use of simply more transmitting BS with $T > R$. As can be seen from the curves in figure 2(a), even subtraction of photons does not allow obtaining squeezing of the quadrature noise less than that of the initial SMSV state. The curves $S_2^{(1)}, S_4^{(1)}$ and $S_6^{(1)}$ can approach the $S_{SMSV}$ curve from above with $S$ growing. The explanation of this fact is the same as above on the basis of observation for $S_{2m}^{(1)}$ with $S \to 0$. The measurement-induced CV states with added single photon and $2m$ subtracted photons are odd, which in the case of $S \to 0$ brings them closer to a single photon whose quadrature uncertainty is more of vacuum one.

The initial discrepancy of the quadrature noises is maintained throughout the entire change of $S$. The corresponding $g_2^{(1)}, g_4^{(1)}, g_6^{(1)}$ take negative values, i.e. $g_2^{(1)}, g_4^{(1)}, g_6^{(1)} < 0$, as shown in Figure 2(c). Conversely, the measurement-induced CV states with odd number $k = 3,5$ of subtracted photons can exhibit squeezing greater than that in the original SMSV state. This is evident when observe the dependencies $g_k^{(1)}$ on $S$ in figure 2(c) which, in particular, shows areas, where the gain of the squeezing $g_3^{(1)}, g_5^{(1)} > 0$ is observed. In the region, where $g_3^{(1)}, g_5^{(1)} < 0$, generated squeezed light has more quadrature noisy component compared to the original SMSV state. As in the case of the vacuum state in the auxiliary mode, the maximum squeezing gain is observed in the CV states with larger number of subtracted photons, that is, an inequality $g_{5,max}^{(1)} > g_{3,max}^{(1)}$ holds. But the width of the squeezing gain $g_3^{(1)}$ is greater than $g_5^{(1)}$ and can be estimated as $S < 4.9\ dB$.

As for the CV state with single photon added and single photon subtracted

$$\left|\Psi_1^{(1)}(y_1, B)\right\rangle = \frac{1}{\sqrt{G_1^{(1)}(y_1,B)}} \sum_{n=0}^{\infty} \frac{y_1^n}{\sqrt{(2n)!}} \frac{(2n)!}{n!}(1 - 2Bn)|2n\rangle \qquad (12)$$

its squeezing is almost comparable to one of the SMSV state, i.e. $S_1^{(1)} \approx S_{SMSV}$ (the corresponding curves almost coincide in the figure 2(a)), which also corresponds to almost horizontal line of the gain of the squeezing $g_1^{(1)} \cong 1$ in figure 2(c). The state in equation (12) can be of interest since it turns into the original SMSV state when $B \to 0$. In addition, the probability of generating the CV state tends to unity $P_1^{(1)} \to 1$ with $B \to 0$ as shown in the figure 2(d), which corresponds to the case of the passage of the SMSV state and a single photon through the highly transmitting BS without reflection into the neighboring modes. Other probabilities $P_2^{(1)}, P_3^{(1)}, P_4^{(1)}, P_5^{(1)}, P_6^{(1)}$ may tend to zero under $B \to 0$, but nevertheless, increasing the parameter $B$ allows the remaining probabilities to improve, especially significantly $P_2^{(1)}$ with $S$ growing.

Another CV state with single photon added and three photons subtracted

$$\left|\Psi_3^{(1)}(y_1, B)\right\rangle = \frac{1}{\sqrt{G_3^{(1)}(y_1,B)}} \sum_{n=0}^{\infty} \frac{y_1^n}{\sqrt{(2n)!}} \frac{(2(n+1))!}{(n+1)!}\left(1 - \frac{2B}{3}n\right)|2n\rangle, \qquad (13)$$

in addition to that it may have a less noisy quadrature component, it has higher average number of photons on compared with original SMSV state $\langle n_3^{(1)} \rangle > \langle n_{SMSV} \rangle$, as shown in fig. 3(a) almost throughout the entire change of input squeezing $S$. Here the mean number of photons $\langle n_k^{(1)} \rangle$ follows from the expression (10) and graphs in figure 3 are presented for those values of $B = B_{opt,k}^{(1)}$ in figure 2(b) that have already been used when constructing the curves in Figures 2(a,c,d). The graphs in Figure 3(a) are shown for extended range of input squeezing $S$ to show the possibility of increasing the brightness of the generated CV state in equation (13) with increase of $S$. The difference in brightness, that is, the ratio $\langle n_3^{(1)} \rangle / \langle n_{SMSV} \rangle$ can reach values greater than 4 just for those $S$ for which the even CV state in equation (13) has less quadrature noise compared to the initial SMSV state which indicates the generation of brighter and more squeezed CV state compared to the original one. In addition, we also show dependencies $\langle n_{SMSV} \rangle$ and $\langle n_1^{(1)} \rangle$ which coincide as well as $\langle n_2^{(1)} \rangle$ which exceeds $\langle n_{SMSV} \rangle$ for all $S$ values. The increased brightness of the generated CV state in Eq. (13) can be explained by its distribution $P_{k,n}^{(1)}$ over Fock states presented in figure 3(b), where the distributions $P_{SMSV,n}$ (SMSV state) and $P_{1,n}^{(1)}$ (CV state in equation (12)) are also presented for comparison which almost coincide $P_{SMSV,n} \cong P_{1,n}^{(1)}$, but the coincidence occurs due to the selected condition $B_{opt,1}^{(1)} \to 0$. But the coincidence of distributions disappears with increasing $B_{opt,1}^{(1)}$. If in SMSV and $|\Psi_1^{(1)}\rangle$ states the vacuum contribution dominates, then in the $|\Psi_3^{(1)}\rangle$ state a shift to multiphoton states occurs, which guarantees increased brightness of the state. As for the probability of success $P_3^{(1)}$ of its generation, it can be significantly increased with increase of the value of $B$.

## 4. Gain of squeezing achieved by imperfect PNR detector

Above we considered the possibility of reducing quadrature noise of the SMSV state from which an even number of photons are subtracted, a procedure that is implemented by an ideal PNR detector, that is, a detector whose quantum efficiency $\eta = 1$. In practice, the quantum efficiency of the PNR detector is not unit, which indicates in favor of considering the generation of more squeezed states in the practical case of $\eta \neq 1$. To do it we use a measuring approach with positive operator-valued measure (POVM) measurement [37] whose elements can be found in Appendix B and which allows us to estimate in the second order by the parameter $1 - \eta$ the quadrature variance of the CV states generated by the PNR detector with quantum efficiency $\eta$

$$(\Delta X_2^2)_{k,\eta}^{(0)} = \frac{1}{4} + \frac{1}{g_k^{(0)}(y_1,B)} \begin{pmatrix} \frac{\langle n_k^{(0)} \rangle}{2}(1-2y_1) - y_1(k+1) + (1-\eta)B\langle n_k^{(0)} \rangle \\ \left(\frac{\langle n_{k+1}^{(0)} \rangle}{2}(1-2y_1) - y_1(k+2)\right) + \frac{(1-\eta)^2}{2!} \\ B^2\langle n_k^{(0)} \rangle \langle n_{k+1}^{(0)} \rangle \left(\frac{\langle n_{k+2}^{(0)} \rangle}{2}(1-2y_1) - y_1(k+3)\right) \end{pmatrix}, \quad (14)$$

where

$$g_{2m}^{(0)}(y_1,B) = 1 + (1-\eta)B\langle n_k^{(0)} \rangle + \frac{(1-\eta)^2}{2!} B^2 \langle n_k^{(0)} \rangle \langle n_{k+1}^{(0)} \rangle. \quad (15)$$

Success probability to generate the measurement-induced CV state by PNR detector with quantum efficiency $\eta$ is given by

$$P_{k,\eta}^{(0)}(\eta) = \eta^k P_k^{(0)} g_{2m}^{(0)}(y_1,B), \quad (16)$$

where formula for $P_k^{(0)}$ is present in Eq. (5).

Plots in figure 4(a-d) show the characteristics of the CV states realized by the PNR detector with quantum efficiency $\eta = 0.6$. So in figures 4(a) the quadrature uncertainties in equation (14) of even CV states in Eq. (B2) optimized by values $B$, i.e., $S_{k,\eta}^{(0)}$, in dependence on input squeezing $S$ are shown. It is interesting to note that optimizing $B_{opt,k,\eta}^{(0)}$ in Figure 4(b) take on almost the same values for all $k = 2,4,6$ used, i.e. $B_{opt,2,0.6}^{(0)} \approx B_{opt,4,0.6}^{(0)} \approx B_{opt,6,0.6}^{(0)} \approx 0.0203$, which is in favor of using only highly transmitting BS together with PNR detector with $\eta = 0.6$ unlike the case of $\eta = 1$. The gain of the squeezing $S_{SMSV} > S_{k,\eta}^{(0)}$ or the same $g_{6,\eta}^{(0)} > 0$ is observed in figure 4(c) in all cases of photon subtraction $k = 2,4,6$ investigated, despite the possibility for the PNR detector to make a mistake in discriminating the measurement outcomes with the maximum being realized for $k = 6$ with $S = 0.86 \, dB$. In general, the condition $g_{6,\eta=0.6,max}^{(0)} > g_{4,\eta=0.6,max}^{(0)} > g_{2,\eta=0.6,max}^{(0)}$ is met. The width of the squeezing gain related to two-photon subtraction can be estimated at $S = 4.7 \, dB$ which is approximately $0.3 \, dB$ less than the width of the gain of squeezing in the case of using a perfect PNR detector. It is interesting to note that the width of the squeezing gain for $g_{2,\eta=0.6}^{(0)}$ can be partly estimated to be twice as large as for the functions $g_{4,\eta=0.6}^{(0)}$ and $g_{6,\eta=0.6}^{(0)}$, in contrast to the case of the CV generation with perfect PNR detector. In addition, as shown in Fig. 4(d), the probability $P_{2,\eta=0.6}^{(0)}$ of subtracting two photons from the SMSV state prevails over the others, i.e. $P_{2,\eta=0.6}^{(0)} > P_{4,\eta=0.6}^{(0)}, P_{6,\eta=0.6}^{(0)}$, especially with increasing $S$. Note that the probability of measuring photons with an imperfect PNR detector decreases compared to ideal detection, which is due to the fact that the detector with $\eta \neq 1$ may no longer register a significant portion of incoming $k$ photons. In general, although the maximum squeezing gain $g_{6,\eta=0.6,max}^{(0)}$ can be achieved by subtracting 6 photons, the strategy of subtracting two photons is preferable even with use of the imperfect PNR detector due to the greater gain width, decrease of the optimized quadrature variance $S_{2,\eta}^{(0)}$ in a wide range of initial $S$ and the increased success probability $P_{2,\eta}^{(0)}$. It is interesting to note that the maximum values of the squeezing $S_{2,\eta}^{(0)}$ are turned out to be comparable with those that could be realized by a perfect PNR detector.

## 5. Conclusion

We analyzed the possibility of obtaining more squeezed light compared to the original SMSV state by pre-subtracting an even number of photons from it. The model used is based on the direct action of the beam splitter operator on the input SMSV state without using the replacement of the BS operator with annihilation operator raised to the power corresponding to the number of photons subtracted, that is, $BS|\Psi\rangle \sim a^k|\Psi\rangle$, which may be acceptable in the case of using highly transmitting BS. The model with the BS operator allows to accurately find all characteristics of the measurement-induced CV states. Both in terms of the width of the squeezing gain, absolute noise squeezing and the probability to conditionally realize the corresponding state, the subtraction of two photons from the initial SMSV state is turned out to be best strategy. It is interesting to note that the advantages are retained even in the case of using an imperfect PNR detector, in particular with quantum efficiency $\eta = 0.6$. Although it is worth noting that the parameters may decrease slightly in the case of using the imperfect PNR detector. One can say that the width of the squeezing gain in the region $< 5\ dB$ is stable with respect to the quantum efficiency of the PNR detector registering two photons. However, in the case of subtracting 4 and 6 photons, it is possible to obtain maximum squeezing gain more than $3\ dB$ with both a perfect and an imperfect PNR detector, but only for small values of initial squeezing. The final noise squeezing of CV states of a certain parity can be calculated in $dB$ as $S_k^{(0,1)}(dB) = g_k^{(0,1)}(dB) + S(dB)$. The squeezing gain of more than $3\ dB$ is greater than that predicted by the approach using annihilation operators to the power corresponding to the number of photons subtracted applied to the original state. Moreover, unlike reference [42], the substantial amplification of the squeezing does not require additional Gaussification procedure, the gain of the squeezing can be realized by direct subtraction of photons from the original SMSV state. Note that the exact theory also predicts a greater value of squeezing of the initial SMSV state from which two photons are subtracted without further Gaussification procedure [46]. So, if we take the initial squeezing $S = 2.4\ dB$, then after subtracting two photons the final squeezing is equal to $S_2^{(0)} = 4.7\ dB$. If we use the PNR detector with $\eta = 0.6$, then the initial SMSV state with squeezing of $S = 0.86\ dB$ can be converted to the even measurement-induced CV state with squeezing of $S_6^{(0)} = 3.96\ dB$ by subtracting 6 photons.

Adding input photons can be an additional factor contributing to the conditional generation of brighter squeezed CV states. At least adding photon and subtracting three photons allows for one to realize a brighter measurement-induced CV state with less quadrature noise than would be the case in the original SMSV state. Further development of the approach with additional CV states as sources of added photons in the auxiliary mode of BS is of interest.

**Appendix A. Measurement-induced generation of the CV states of definite parity**

For the measurement-induced generation of CV states (3,4) we make use of BS with arbitrary real transmittance $t > 0$ and reflectance $r > 0$ subject to the normalization condition $t^2 + r^2 = 1$ that is described by the matrix $BS_{12} = \begin{bmatrix} t & -r \\ r & t \end{bmatrix}$. The BS mixes the modes 1 and 2 converting input creation operators $a_1^+$ and $a_2^+$ as follows: $BS_{12} a_1^+ BS_{12}^+ = ta_1^+ - ra_2^+$ and $BS_{12} a_2^+ BS_{12}^+ = ra_1^+ + ta_2^+$, respectively. It transforms the original SMSV state in Eq. (1) into a hybrid entangled state [35,37]

$$BS_{12}(|SMSV(y)\rangle_1|0\rangle_2) = \frac{1}{\sqrt{\cosh s}} \sum_{k=0}^{\infty} c_k^{(0)}(y_1, B)\sqrt{Z^{(k)}(y_1)}|\Psi_k^{(0)}(y_1)\rangle_1 |k\rangle_2, \quad (A1)$$

whose amplitudes are given by

$$c_k^{(0)}(y_1, B) = (-1)^k \frac{(y_1 B)^{\frac{k}{2}}}{\sqrt{k!}}, \quad (A2)$$

while CV states of definite parity are represented by formulas (3,4), where the input squeezing parameter $y$ decreases by $t^2$ times, that is, it becomes equal to $y_1 = yt^2 = y/(1 + B) \leq y$ and the BS parameter used is equal to $B = (1 - t^2)/t^2$. The normalization factors of the CV states of a certain parity in Eqs. (3,4) are determined through $2m, 2m + 1$ derivatives of the analytical function $Z(y_1) = 1/\sqrt{1 - 4y_1^2}$, i.e., $Z^{(2m)}(y_1) = dZ^{2m}/dy_1^{2m}$ and $Z^{(2m+1)}(y_1) = dZ^{2m+1}/dy_1^{2m+1}$, respectively.

Adding single photon to original SMSV state can be realized by means of the state in Eq. (A1) which is obtained after the passage of the SMSV state through BS as

$$BS_{12}(|SMSV(y)\rangle_1|1\rangle_2) = \frac{1}{\sqrt{\cosh s}} \sum_{k=0}^{\infty} c_k^{(1)}(y_1, B)\sqrt{G_k^{(1)}(y_1, B)}|\Psi_k^{(1)}(y_1, B)\rangle_1 |k\rangle_2 \quad (A3)$$

with the measurement-induced CV states of definite parity

$$|\Psi_0^{(1)}(y_1)\rangle = \frac{1}{\sqrt{G_0^{(1)}(y_1)}} \sum_{n=0}^{\infty} \frac{y_1^n}{\sqrt{(2n+1)!}} \frac{(2n)!}{(n)!}(2n + 1)|2n + 1\rangle, \quad (A4)$$

$$|\Psi_{2m}^{(1)}(y_1, B)\rangle = \sqrt{\frac{y_1}{G_{2m}^{(1)}(y_1, B)}} \sum_{n=0}^{\infty} \frac{y_1^n}{\sqrt{(2n+1)!}} \frac{(2(n+m))!}{(n+m)!}\left(1 - \frac{2n+1}{2m}B\right)|2n + 1\rangle, \quad (A5)$$

$$|\Psi_{2m+1}^{(1)}(y_1, B)\rangle = \frac{1}{\sqrt{G_{2m+1}^{(1)}(y_1, B)}} \sum_{n=0}^{\infty} \frac{y_1^n}{\sqrt{(2n)!}} \frac{(2(n+m))!}{(n+m)!}\left(1 - \frac{2n}{2m+1}B\right)|2n\rangle, \quad (A6)$$

whose amplitudes are given by

$$c_k^{(1)}(y_1, B) = \frac{1}{\sqrt{1+B}} \begin{cases} \sqrt{B}, & \text{if } k = 0 \\ (-1)^{k+1} \frac{(y_1 B)^{\frac{k-1}{2}}}{\sqrt{k!}} k, & \text{if } k \neq 0 \end{cases}. \quad (A7)$$

As the CV states $|\Psi_k^{(1)}(y_1, B)\rangle$ have an additional internal term in contrast to the CV states without added single photon their normalization factors have more complex form

$$G_0^{(1)}(y_1) = \frac{d}{dy_1}(y_1 Z(y_1)) = Z^3(y_1), \quad (A8)$$

$$G_k^{(1)}(y_1, B) = Z^{(k-1)}(y_1) + a_{k,1}^{(1)}(B)\left(y_1 Z^{(k)}(y_1)\right) + a_{k,2}^{(1)}(B) y_1 \frac{d}{dy_1}\left(y_1 Z^{(k)}(y_1)\right), \quad (A9)$$

where their coefficients are the following

$$a_{k,1}^{(1)}(B) = -\frac{2B}{k}, \quad (A10)$$

$$a_{k,2}^{(1)}(B) = \left(\frac{B}{k}\right)^2. \tag{A11}$$

As in the case discussed above, the subscript $k$ indicates the number of photons being subtracted while the superscript $(1)$ is responsible for the number of photons being added. Using the CV states with single photon added in equations (A4-A6), it is possible to derive analytical form of the functions appearing in expressions for squeezing in equation (9), i.e.

$$R_k^{(1)}(y_1, B) = kZ^{(k-1)}(y_1) + (1-B)\left(y_1 Z^{(k)}(y_1)\right) - \frac{B}{k} y_1 \frac{d}{dy_1}\left(y_1 Z^{(k)}(y_1)\right). \tag{A12}$$

## Appendix B. Output squeezing with imperfect PNR detector

Here we are interested in considering influence of an imperfect PNR detector on the output state. Model of the PNR detector with quantum efficiency $\eta$ can be modeled by means of use POVM formalism with POVM elements $\{\Pi_k, k = 0, \ldots, \infty\}$

$$\Pi_k = \eta^k \sum_{x=0}^{\infty} \binom{C_{2(m+x)}^{2m}(1-\eta)^{2x} + |2(m+x)\rangle\langle 2(m+x)| +}{C_{2(m+x)+1}^{2m}(1-\eta)^{2x+1} + |2(m+x)+1\rangle\langle 2(m+x)+1|}, \tag{B1}$$

where $C_{2(m+x)}^{2m}$ and $C_{2(m+x)+1}^{2m}$ are the binomial coefficients. The case $\eta = 1$ corresponds to a perfect PNR detector, and in the case under consideration, we accept $\eta$ in wide range of its change.

Let us consider the CV state generated by the imperfect PNR detector in the case of the passage of the SMSV state through the BS. Using the general definition of the measurement-induced CV state realized with corresponding measurement operator in equation (B1) $\rho_k^{(0)} = tr_2\left(\rho_{12}^{(0)} \Pi_k\right)/P_k^{(0)}(\eta)$, where the operation $tr_2$ means taking a trace over the states in the second mode, the state $\rho_{12}^{(0)}$ is the one in Eqs. (3,4) rewritten in terms of the density matrix and $\rho_k^{(0)}$ and $P_k^{(0)}(\eta)$ are the corresponding output state and its probability, and taking into account the terms up to the second order of smallness on parameter $\eta$, i.e. $(1-\eta)^2$, finally, we have

$$\rho_k^{(0)} = \frac{1}{g_k^{(0)}} \binom{|\Psi_k^{(0)}(y_1)\rangle\langle\Psi_k^{(0)}(y_1)| + (1-\eta)B\langle n_k\rangle|\Psi_{k+1}^{(0)}(y_1)\rangle\langle\Psi_{k+1}^{(0)}(y_1)| +}{\frac{(1-\eta)^2}{2!}B^2\langle n_k\rangle\langle n_{k+1}\rangle|\Psi_{k+2}^{(0)}(y_1)\rangle\langle\Psi_{k+2}^{(0)}(y_1)|}, \tag{B2}$$

where $g_k^{(0)}$ is given by Eq. (15). Using the formula, it is possible to find characteristics of the generated CV states of a certain parity in equations (3,4) in particular, expression for the quadrature noise in Eq. (14). Note that the CV states already realized by the PNR detector with non-unit quantum efficiency are not CV states of definite parity. As can be seen from the expression in Eq. (B2), the states can already contain both even and odd Fock states, which define them as CV states of indefinite parity.

## Acknowledgments

The work of MSP and SAP was supported by the Foundation for the Advancement of Theoretical Physics and Mathematics "BASIS".

## References

[1] H. P. Yuen, "Two-photon coherent states of the radiation field," Phys. Rev. A **13**, 2226–2243 (1976).
[2] R. Slusher, L. Hollberg, B. Yurke, J. Mertz and J. Valley, "Observation of squeezed states generated by four-wave mixing in an optical cavity," Phys. Rev. Lett. **55**, 2409-2412 (1985).

[3] R. Loudon and P. Knight, "Squeezed light," Journal of Modern Optics **34**, 709–759 (1987).
[4] D.F. Walls, "Squeezed states of light," Nature **306**, 141-146 (1983).
[5] P. Buscha, T. Heinonen and P. Lahti, "Heisenberg's uncertainty principle," Physics Reports **452**, 155-176 (2007).
[6] B.L. Schumaker and C.M. Caves, New formalism for two-photon quantum optics. II. Mathematical foundation and compact notation," Phys. Rev. A **31**, 3093-3111 (1985).
[7] C.C. Gerry and P.L. Knight, Introductory quantum optics, Cambridge University Press, 2005.
[8] L.A. Wu, H.J. Kimble, J.L. Hall and H. Wu, "Generation of squeezed states by parametric down conversion," Phys. Rev. Lett. **57**, 2520-2523 (1986).
[9] P. Kürz, R. Paschotta, K. Fiedler, A. Sizmann, G. Leuchs and J. Mlynek, "Squeezing by second-harmonic generation in a monolithic resonator," Appl. Phys. B **55**, 216-225 (1992).
[10] K. Schneider, M. Lang and S. Schiller, "Generation of strongly squeezed continuous-wave light at 1064 nm," Opt. Express **2**, 59-64 (1998).
[11] J.W. Wu, P.K. Lam, M.B. Gray and H.-A. Bachor, "Optical tomography of information carrying laser beams," Opt. Express **3**, 154-161 (1998).
[12] Y. Takeno, M Yukawa, H, Yonezawa and A. Furusawa, "Observation of -9 dB quadrature squeezing with improvement of phase stability in homodyne measurement," Opt. Express, **15** 4321-4327 (2007).
[13] H. Vahlbruch, M, Mehmet, S. Chelkowski, B. Hage, A. Franzen, N. Lastzka, S. Goßler, K. Danzmann and R. Schnabel, "Observation of squeezed light with 10-dB quantum noise reduction," Phys. Rev. Lett. **100**, 033602 (2008).
[14] M. Mehmet, S. Ast, T. Eberle, S. Steinlechner, H. Vahlbruch and R. Schnabel, "Squeezed light at 1550 nm with a quantum noise reduction of 12.3 dB," Opt. Express **19**, 25763-25772 (2011).
[15] H. Vahlbruch, M. Mehmet, K. Danzmann and R. Schnabel, "Detection of 15 dB squeezed states of light and their application for the absolute calibration of photoelectric quantum efficiency," Phys. Rev. Lett. **117**, 110801 (2016).
[16] C.M. Caves, "Quantum-mechanical noise in an interferometer", Phys. Rev. D **23** 1693-1708 (1981).
[17] L. Pezze and A. Smerzi, "Mach-Zehnder interferometry at the Heisenberg limit with coherent and squeezed-vacuum light," Phys. Rev. Lett. **100**, 073601 (2008).
[18] J. Abadie *et al.*, "A gravitational wave observatory operating beyond the quantum shot-noise limit," Nature Physics, **7**, 962-965 (2011).
[19] J. Aasi *et al.*, "Enhanced sensitivity of LIGO gravitational wave detector by using squeezed states of light," Nature Photonics **7**, 613-619 (2013).
[20] F. Acernese *et al.*, "Increasing the astrophysical reach of the advanced Vigro detector via the application of squeezed vacuum states of light," Phys. Rev. Lett. **123**, 231108 (2019).
[21] R. Schnabel, "Squeezed states of light and their applications in laser interferometers," Physics Reports **684** 1-51 (2017).
[22] A. Furusawa, J.L. Sörensen, S.L. Braunstein, C.A. Fuchs, H.J. Kimble and E.S. Polsik, "Unconditional quantum teleportation," Science **282**, 706-709 (1998).
[23] H. Yonezawa, S.L. Braunstein and A. Furusawa, "Experimental demonstration of quantum teleportation of broadband squeezing," Phys. Rev. Lett. **99**, 110503 (2007).
[24] S. Takeda, T. Miuta, M. Fuwa, P. van Loock and A. Furusawa, "Deterministic quantum teleportation of photonic quantum bits by a hybrid technique," Nature **500**, 315-318 (2013).
[25] A.I. Lvovsky, H. Hansen, T. Aichele, O. Benson, J Mlynek and S. Schiller, "Quantum state reconstruction of the single-photon Fock state," Phys. Rev. Lett. **87**, 050402 (2001).

[26] M. Cooper, L.J. Wright, C. Söller and B.J. Smith, "Experimental generation of multi-photon Fock states," Opt. Express **21**, 5309-5317 (2013).
[27] A. Ourjoumtsev, R. Tualle-Brouri, J. Laurat and P. Grangier, "Generating optical Schrödinger kittens for quantum information processing," Science **312**, 83-86 (2006).
[28] J.S. Neergaard-Nielsen, B.M. Nielsen, C. Hettich, K. Mølmer and E.S. Polzik, "Generation of a superposition of odd photon number states for quantum information networks," Phys. Rev. Lett. **97**, 083604 (2006).
[29] S.A. Podoshvedov, "Building of one-way Hadamard gate for squeezed coherent states," Phys. Rev. A **87**, 012307 (2013).
[30] A. Zavatta, S. Viciani and M. Bellini, "Quantum-to-classical transition with single-photon-added coherent states of light," Science **306**, 660-662 (2004).
[31] J. Wenger, R. Tualle-Brouri and P. Grangier, "Non-gaussian statistics from individual pulses of squeezed light," Phys. Rev. Lett. **92**, 153601 (2004).
[32] T. Gerrits, S. Glancy, T.C. Clement, B, Calkins, A.E. Lita, A.J. Miller, A.L. Migdall, S.W. Nam, R.P. Mirin and E. Knill, "Generation of optical coherent-state superpositions by number-resolved photon subtraction from the squeezed vacuum," Phys. Rev A **82**, 031802 (2010).
[33] M.S. Kim, "Recent developments in photon-level operations on travelling light fields," J. Phys. B **41**, 133001 (2008).
[34] V. Parigi, A. Zavatta, M.S. Kim and M. Bellini, "Probing quantum communication rules by addition and subtraction of single photons to/from a light field," Science **317**, 1890-1893 (2007).
[35] D.A. Kuts, M.S. Podoshvedov, Ba An Nguyen, S.A. Podoshvedov, "A realistic conversion of single-mode squeezed vacuum state to large-amplitude high-fidelity Schrödinger cat states by inefficient photon number resolving detection," Phys. Scr. **97**, 115002 (2022).
[36] C. Kumar, Rishabh, M. Sharma and S. Arora, "Parity-detection-based Mach-Zehnder interferometry with coherent and non-Gaussian squeezed vacuum states as inputs," Phys. Rev. A **108**, 012605 (2023).
[37] M.S. Podoshvedov, S.A. Podoshvedov and S.P. Kulik, "Algorithm of quantum engineering of large-amplitude high-fidelity Schrödinger cat states," Sci. Rep. **13**, 3965 (2023).
[38] D. Fukuda *et al.*, "Titanium-based transition-edge photon number resolving detector with 98% detection efficiency with index-matched small-gap fiber coupling," Opt. Express **19**, 870-875 (2011).
[39] M. Eaton, *et al.*, "Resolution of 100 photons and quantum generation of unbiased random numbers," Nature Photonics **17**, 106-111 (2023).
[40] A. Biswas and G.S. Agarwal, "Nonclassicality and decoherence of photon-subtracted squeezed states," Phys. Rev. A **75**, 032104 (2007).
[41] M.S. Podoshvedov and S.A. Podoshvedov, "Family of CV state of definite parity and their metrological power," **20**, 045202 (2023).
[42] S. Grebien, J. Göttsch, B. Hage, J. Fiurášek and R. Schnabel, "Multistep two-copy distillation of squeezed states via two-photon subtraction," Phys. Rev. Lett. **129**, 273604 (2022).
[43] S.A. Podoshvedov and Nguyen Ba An, Quantum Inf. Process. "Design of interactions between discrete-and continuous-variable states for generation of hybrid entanglement," **18**, 68 (2019).
[44] K. Huang, H. Le Jeannic, O. Morin, T. Darras. G. Guccione, A. Cavaillès and J. Laurat, "Engineering optical hybrid entanglement between discrete-and continuous variable states," New Journal of Phys. **21**, 083033 (2019).

[45] S.A. Podoshvedov and M.S. Podoshvedov, JOSA B "Entanglement synthesis based on the interference of a single-mode squeezed vacuum and delocalized photon," **18**, 1341-1349 (2021).
[46] J. Eisert, D. Browne, S. Schell and M. Plenio, "Distillation of continuous-variable entanglement with optical means," Ann. Phys. (Amsterdam) **311**, 431 (2004).

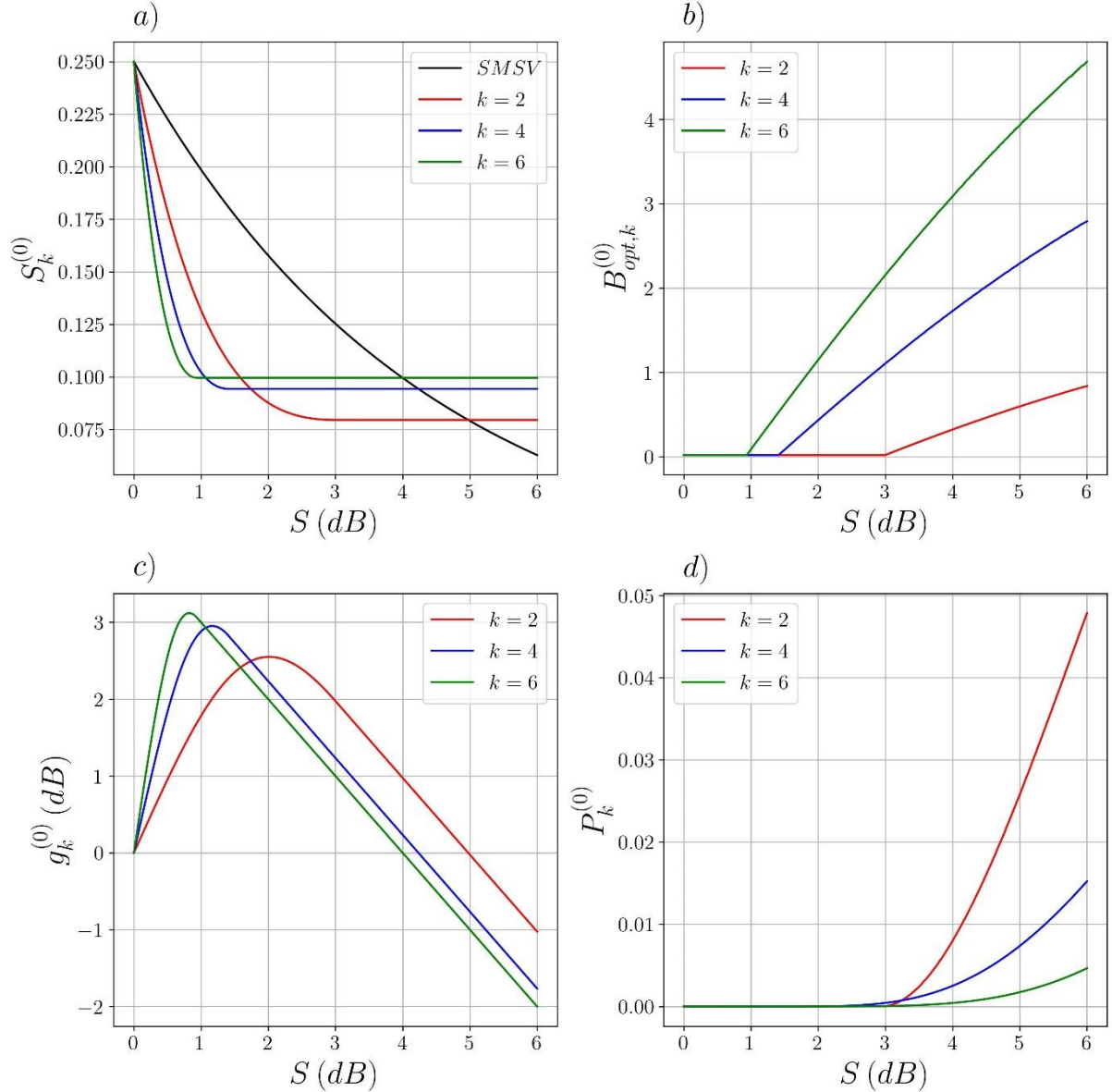

**Fig. 1(a-d).** (a) Dependences of the optimized quadrature variance demonstrated by the measurement-induced even CV states in Eq. (3) on the input squeezing $S$, where values of $B = B_{opt,k}^{(0)}$ which provide minimal values of $S_k^{(0)}$ in (a) for given $S$ and $k = 2,4,6$ are shown in (b). Function $S_{SMSV}$ is presented for comparison. In general, dependencies $B_{opt,k}^{(0)}$ have a fairly sharp transition from regime $B < 1$ (highly transmitting BS) to $B > 1$ (more reflective BS) through $B = 1$ (balanced BS) and it occurs almost linearly starting from some value of $S$. The observed almost horizontal sections in (a) are associated with the corresponding behavior of $B_{opt,k}^{(0)}$. The dependences of the squeezing gain $g_k^{(0)}$ on $S$ are shown in (c). The probabilities $P_k^{(0)}$ of generating even CV states can grow with increase of $S$ as in (d).

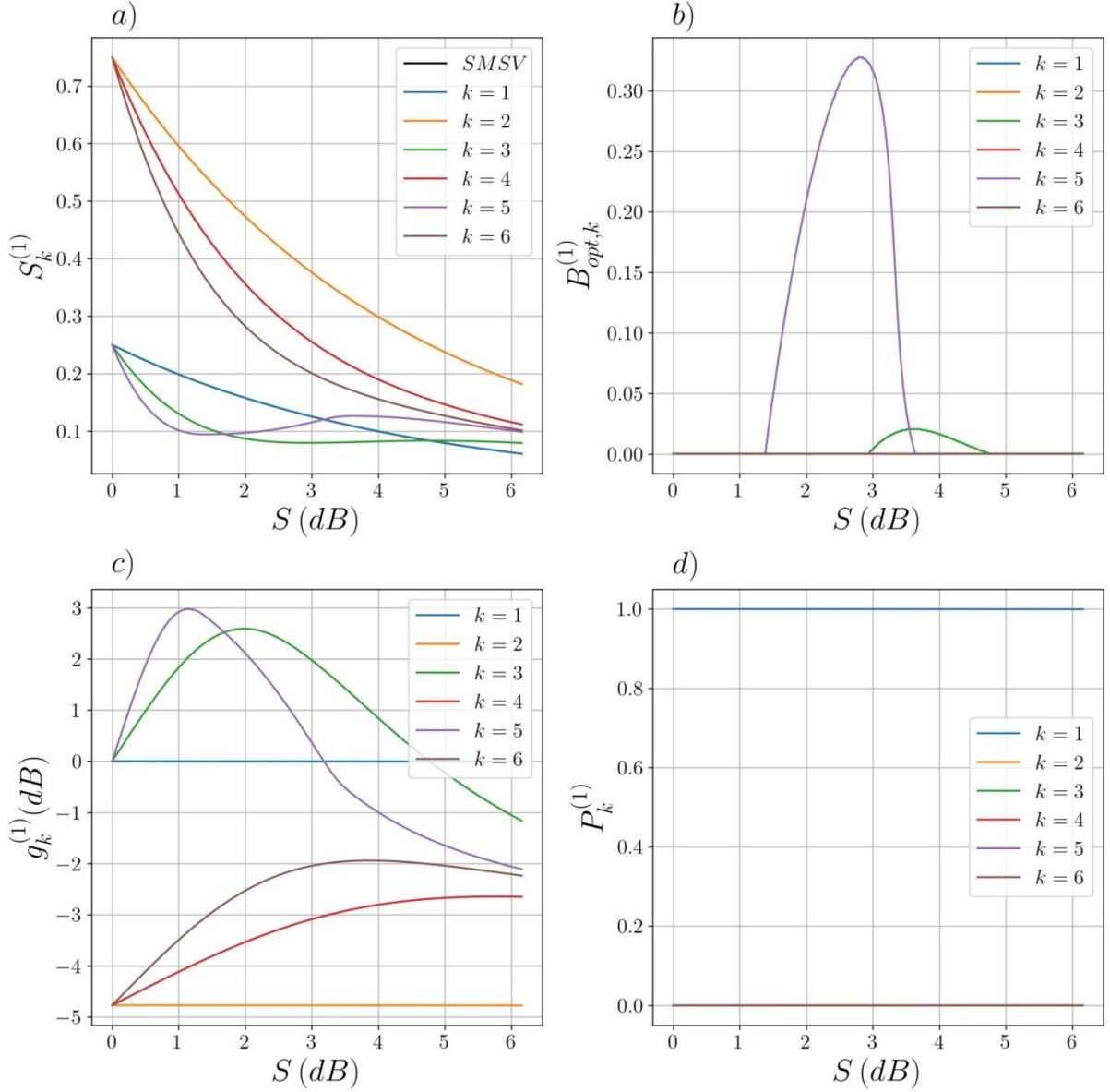

**Fig. 2(a-d).** (a) Dependences of the optimized quadrature variance $S_k^{(1)}$ of the CV states of definite parity generated by an ideal PNR detector in the case of added single photon and $k$ subtracted photons involving $S_{SMSV}$ from the initial squeezing $S$. (b) Dependence of the optimizing values $B_{opt,k}^{(1)}$ which provide $S_k^{(1)}$ in (a) on the same $S$. Dependencies $B_{opt,3}^{(1)}$ and $B_{opt,5}^{(1)}$ can have a bell-shaped form in a certain area of $S$. Plots in (c) of the functions $g_k^{(1)}$ on $S$ allow for one to quantify the gain of the squeezing. Dependence of the success probabilities $P_k^{(1)}$ of generating CV states with addition of a single photon and subtraction of $k$ photons on $S$ in (d) shows that $P_1^{(1)}$ can dominate over all others but only in the case of $B_{opt,k}^{(1)} \to 0$.

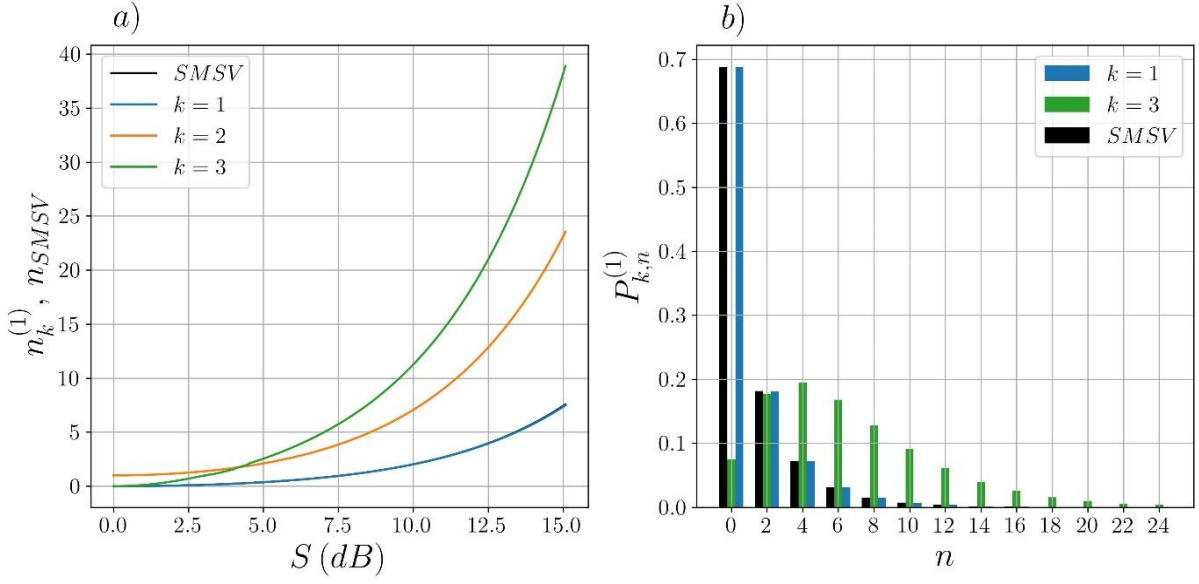

**Fig. 3(a,b).** (a) Dependence of the mean number photons $\langle n_{SMSV}\rangle$ and $\langle n_k^{(1)}\rangle$ for $k = 1,2,3$ on input squeezing $S$ expanded to $15\ dB$. With used value of $B = B_{opt,k}^{(1)}$, $\langle n_{SMSV}\rangle$ and $\langle n_1^{(1)}\rangle$ are the same. The more photons are subtracted, the brighter the generated CV state becomes. (b) Distributions $P_{k,n}^{(1)}$ of the CV states: SMSV, $|\Psi_1^{(1)}\rangle$ and $|\Psi_3^{(1)}\rangle$ over Fock states are in (b). The distribution $P_{k,3}^{(1)}$ in the case of subtraction of three photons $k = 3$ shifts towards multiphoton states in contrast to the coinciding ones $P_{SMSV,n} \approx P_{1,n}^{(1)}$ in which the contribution of the vacuum state prevails. The graphs are obtained for the $B_{opt,k}^{(1)}$ in Fig. 2(b) that are already used to create the curves in the figures 2(a,c,d). The photon distribution $P_{1,n}^{(1)}$ can also shift towards multiphoton states in the case of a choice of $B$ different from the $B_{opt,1}^{(1)}$.

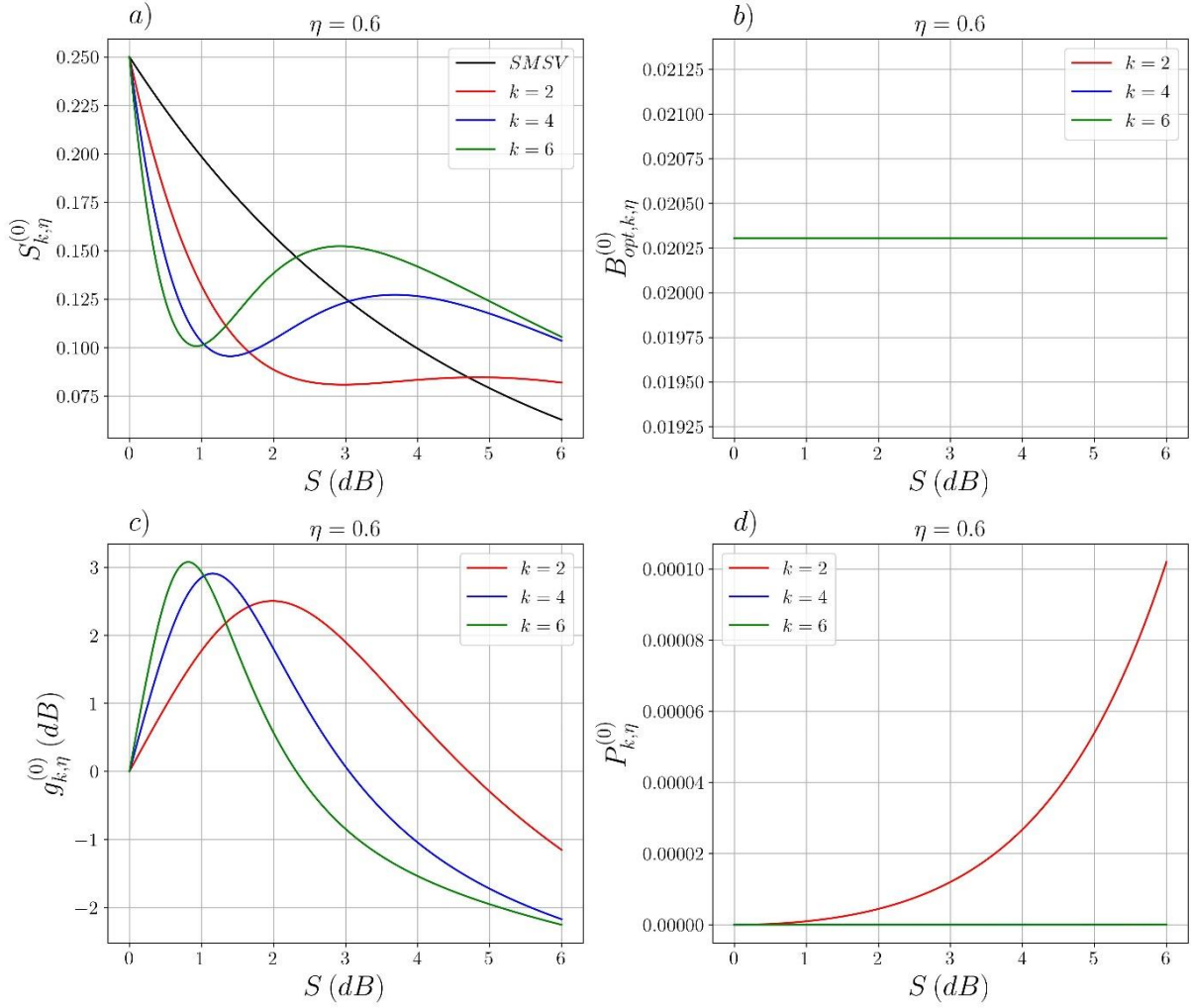

**Fig. 4(a-d).** The dependences of the quadrature uncertainties in equation (14) (a) optimized by $B = B_{opt,k,\eta}^{(0)}$ (b), i.e. $S_{k,\eta}^{(0)}$, on the initial squeezing $S$ are shown in the case of using the PNR detection with quantum efficiency $\eta = 0.6$. Unlike the case of $\eta = 1$, the values of $B_{opt,k,\eta=0.6}^{(0)}$ take on almost identical values which is expressed in the observation of almost horizontal line for all $k = 2,4,6$. The difference between $B_{opt,k,\eta=0.6}^{(0)}$ exists but in much more distant digits after the decimal point. Function $g_{2,\eta=0.6}^{(0)}$ shows a larger width of the gain of the squeezing in (c) compared to $g_{4,\eta=0.6}^{(0)}$ and $g_{6,\eta=0.6}^{(0)}$. The probability of generating the two-photon subtracted CV state $P_{2,\eta=0.6}^{(0)}$ may dominate over other probabilities $P_{4,\eta=0.6}^{(0)}$ and $P_{6,\eta=0.6}^{(0)}$ as shown in (d).